\documentclass[apj]{emulateapj}
\usepackage{xcolor}

\begin{document}

\title{Periodic Eruptive Variability of the Isolated Pre-Main Sequence Star V347 Aurigae}

\author{S. E. Dahm\altaffilmark{1} \& L. A. Hillenbrand\altaffilmark{2}}

\altaffiltext{1}{U.S. Naval Observatory, Flagstaff Station, 10391 West Naval Observatory Road, Flagstaff, AZ 86005-8521, USA}
\altaffiltext{2}{Department of Astronomy, California Institute of Technology, Pasadena, CA 91125, USA}

\begin{abstract}
V347 Aurigae is associated with the small dark cloud L1438 and appears to be an isolated pre-main sequence star located at distance $d\approx$200 pc. Multi-epoch, archival photometry reveals periodic brightness variations with amplitude $V\approx2.0$ magnitudes occurring on timescales of $\sim$160 days that have persisted for decades. Regular cadence, optical imaging of the source with the {\it Zwicky Transient Facility} shows that a small reflection nebula illuminated by V347 Aur also fluctuates in brightness, at times fading completely. Multi-epoch, {\it Keck}/HIRES data suggests the presence of two distinct spectral components: a prominent emission-line dominated spectrum with a heavily veiled continuum correlated with the bright photometric state, and an M-type absorption line spectrum associated with quiescence. All spectra exhibit strong Balmer and He I line emission, consistent with accretion, as well as high velocity emission arising from the forbidden transitions of [O I], [N II], and [S II] that are generally associated with collimated jets and disk winds. There is no evidence in existing high dispersion spectroscopy or high resolution imaging for binarity of V347 Aur. The repeating outburst events are possibly linked to accretion instabilities induced by an undetected companion or a structure within the circumstellar disk that periodically increases the mass accretion rate. V347 Aur is perhaps analogous to an EXor-type variable, though more regularly recurring.
\end{abstract}


\section{Introduction}

The vast majority of stars in the Milky Way form in unbound clusters or associations
that disperse over timescales of $<$10$^{8}$ years. The stellar populations of these 
clusters and associations vary from several hundred members (e.g. IC\,348, NGC\,1333), 
to several thousand (e.g. the Orion Nebula Cluster, the Upper Scorpius OB association). 
Even in dispersed star forming regions, such as the Taurus-Auriga complex or the TW Hydra 
association, hundreds down to dozens of stars have formed together within a multitude of
dense molecular cloud cores, often bridged by lower density molecular gas. Some stars, 
however, form in more isolated environments.

V347 Aurigae (HBC 428) is a highly variable, spectral type M2, luminosity class V 
(Cohen, 1978) star that is still embedded within its natal molecular cloud and appears 
to have formed in relative isolation, far from known star forming regions. V347 Aur lies 
within a small molecular core, Lynds' dark nebula (LDN) 1438 (Lynds 1962). L1438 is one 
of perhaps a dozen small molecular clouds located near the intersection of the constellations 
Auriga, Perseus, and Camelopardalis (Reipurth 2008). The large scale CO (1--0) surveys of 
Dame et al. (2001) show that V347 Aur and its host cloud lie near the edge of this larger 
complex, which is centered roughly $\sim$0.5 degree to the northwest. The radial velocity 
of L1438 measured by Dame et al. (2001) is +5.6 km s$^{-1}$ in the local standard of rest
(LSR), consistent with membership in the local spiral arm. 

Dobashi et al. (2005) applied standard star counting techniques to the digitized sky 
survey and mapped an extensive, isolated dust cloud coincident with the molecular complex 
of Dame et al. (2001). The dust cloud enveloping V347 Aur is identified as number 1046, 
clump P1 and exhibits a maximum extinction of $A_{V}\sim$2.8--3.1 magnitudes (Dobashi 
et al. 2005). 

The pre-main sequence star is clearly interacting with nebular material and is associated 
with a small reflection nebula. Cohen (1978) and Gyul'budagyan \& Magakyan (1977) independently 
discovered the small reflection nebula associated with V347 Aur during a systematic search of 
the Palomar Sky Survey plates for cometary nebulae. Cohen (1978) identified it on two pairs of 
images taken in 1953 October and 1954 December. Cohen (1978) further noted that the star and the 
faint arc of nebulosity associated with it were visible on the 1953 plates, but that both had 
faded considerably by 1954. 

\begin{figure}
\includegraphics[angle=-90,origin=c, height=7cm]{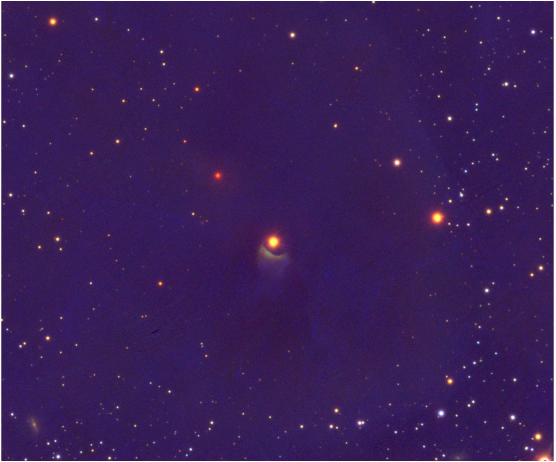}
\caption{Composite, three-color $riy-$band image of the V347 Aur region obtained from the 
Pan-STARRS image server oriented such that north is up and east is to the left. The field 
of view presented here is approximately 8\farcm0$\times$8\farcm0. V347 Aur and its reflection 
nebula lie roughly centered in the L1438 molecular cloud, and several heavily reddened stars 
lie along the cloud periphery.}
\label{fig:psimage}
\end{figure}

V347 Aur was included as object number 33 in Cohen's (1980) catalog of red nebulous objects (RNO) 
associated with dark nebulae. The reflection nebula illuminated by V347 Aur is shown in 
Figure~\ref{fig:psimage} , a composite, $riy-$band image of the region obtained from the Panoramic 
Survey Telescope and Rapid Response System (Pan-STARRS). L1438 is readily apparent as a small dark 
nebula, perforated in its center by V347 Aur. Several heavily reddened stars are evident along the 
cloud periphery with maximum extinction occurring near its center, where V347 Aur has formed.

In this contribution, we first provide a summary of previous knowledge regarding the stellar 
properties of the young pre-main sequence object V347 Aur. Then, archival photometry obtained 
from multiple, publicly available sources including the Zwicky Transient Facility (ZTF), the 
Palomar Transient Factory (PTF), the Northern Sky Variability Survey (NSVS), and the All-Sky 
Automated Survey for Supernovae (ASASSN) are examined to quantify the amplitude and periodicity 
of the V347 Aur brightness variations. Archival and new optical, high dispersion spectra of 
V347 Aur obtained using the {\it Keck I} telescope are presented, which reveal significant 
changes in the spectrum over timescales of months to years. A high dispersion, near infrared 
spectrum of the source from the {\it Keck II} telescope is also presented, showing structural 
details of the \ion{He}{1} 10830 and Paschen $\gamma$ emission lines. Archival Hubble Space 
Telescope (HST) imaging of V347 Aur is examined for spatially resolved companions. Finally, 
the source of variability and the isolated environment of the pre-main sequence object are 
discussed in more detail. 

\section{Properties of V347 Aur}

The variability of V347 Aur was first recognized by Morgenroth (1939), who gave a photographic 
magnitude range of $m_{pg}$=14--$>$15.5 mag and suggested it to be a possible long-period variable. 
Wenzel (1978) examined 150 blue sensitive plates taken with a 14 cm camera, finding V347 Aur to 
be invisible on most (implying a photographic magnitude fainter than $m_{pg}\sim$15.3 to 15.8). 
On four occasions, however, it was observed to be well above the plate limit: in 1931 November 
(13.5 mag), 1934 November (13.1 mag), 1945 March (14.4 mag), and 1967 December (14.7 mag). 
Wenzel (1978) concluded that a 346-day period would fit these observed maxima.

Cohen (1978) obtained a low dispersion, optical scanner spectrogram of V347 Aur in 1977, describing 
it as an emission line spectrum, recognizably that of a T Tauri star, overlaid upon an absorption 
line spectrum. He classified V347 Aur as spectral type M2, probably luminosity class V based upon 
the strengths of the CaH and TiO bandheads at $\lambda\lambda$4762, 4955, 5440, and 6159. The shape 
of the continuum implied an extinction of $A_{V}$=2.20$\pm$0.40 mag, whereas the Balmer decrement 
suggested a substantially larger extinction of $A_{V}$=5.02$\pm$0.10 mag. Given its emission line 
spectrum and placement on the color-magnitude diagram, Cohen (1978) concluded the star to be extremely young.

Modern photometric studies of V347 Aur begin with Wozniak et al. (2004), who included the object in their catalog of slowly varying stars with near infrared colors consistent with the asymptotic giant branch population (object ID 4314476).  They identified V347 Aur as a Mira-type variable and carbon star.

Connelley \& Greene (2010) obtained a moderate dispersion, near infrared spectrum of V347 Aur using SpeX on the NASA Infrared Telescope Facility (IRTF) in 2007. These authors classified the source as an M2$\pm$1 spectral type.  The peaked $H-$band spectral profile of V347 Aur suggested a low surface gravity, consistent with its pre-main sequence nature. Connelley \& Greene (2010) also noted significant extinction: $A_{V}\sim$3.4 mag (determined from its near infrared colors), 4.4 mag (determined by fitting the continuum with the best matching spectral type), and 7.5 mag 
(from the Pa$\beta$--Br$\gamma$ line ratio analysis). The spectral index determined by Connelley \& Greene (2010), $\alpha$=0.05,  implies a flat-spectrum source, i.e. a transitional Class I protostar -- Class II Classical T Tauri star (Lada \& Wilking 1984).

Flores et al. (2019) used the cross-dispersed, high resolution, near infrared spectrograph iSHELL on the IRTF to measure the magnetic field strength associated with V347 Aur and derived several fundamental properties. Their best fit model parameters included effective temperature, T$_{e}$=3233 K; log($g$)=3.25; vsin($i$)=11.7 km s$^{-1}$; and magnetic field strength $B$=1.36 kG. Flores et al. (2019) placed V347 Aur on the color-magnitude diagram to compare its position with the evolutionary models of Baraffe et al. (2015) and Feiden (2016), finding a track-dependent mass in the range $\sim$0.2--0.4 M$_{\odot}$ and an age $<$1 Myr.

\begin{figure}
\epsscale{1}
\plotone{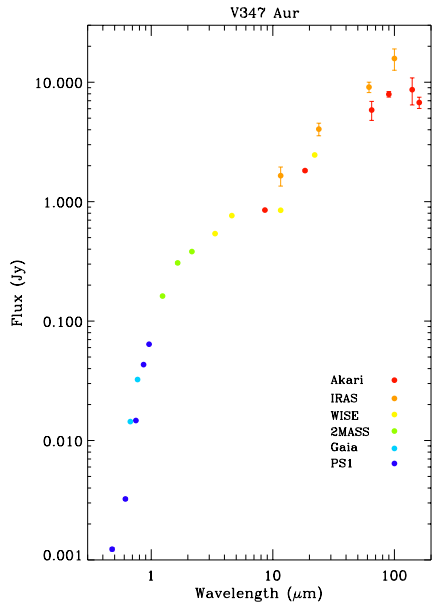}
\caption{The spectral energy distribution of V347 Aur reproduced from Pan-STARRS, {\it Gaia}, 2MASS, WISE, AKARI, and IRAS photometry demonstrating the presence of substantial circumstellar material.
}
\label{fig:sed}
\end{figure}

The presence of substantial circumstellar material associated with V347 Aur is inferred from its spectral energy distribution (SED), which is reproduced in Figure~\ref{fig:sed} using optical, near and mid-infrared photometry from Pan-STARRS, {\it Gaia}, 2MASS, the Widefield Infrared Survey Explorer (WISE) and far infrared photometry from the Infrared Astronomical Satellite (IRAS) and AKARI. The steep rise in the SED beyond $\sim$10$\mu$m results from the reprocessing of starlight by circumstellar dust in the outer disk and envelope.

\section{New Observations}

High-dispersion, optical spectra of V347 Aur were obtained on the nights of 2003 December 13, 
2004 September 24, 2004 November 21, and 2005 November 23 by G. H. Herbig, and on the night of 
2019 November 29 by LAH using the High Resolution Echelle Spectrometer (HIRES), Vogt et al. (1994), 
installed at the Nasmyth focus of the {\it Keck I} telescope. The 2003 observation by Herbig was 
made using the single Tektronix detector, while all later observations were made with the upgraded 
MIT/Lincoln Laboratory three-chip mosaic. HIRES was configured with the red cross-disperser and 
collimator in beam. The C1 decker or slit (0\farcs87$\times$7\farcs) was used for the earlier data, 
providing a spectral resolution of $\sim$45,000 ($\sim$6.7 km s$^{-1}$) while the C5 decker 
(1\farcs14$\times$7\farcs) was used for the 2019 observation resulting in $\sim$34,000 
($\sim$8.9 km s$^{-1}$) spectra. Near-complete wavelength coverage from $\sim$4300 
to 8600 \AA\ was achieved for the 2004, 2005, and 2019 observations, whereas the 2003 observation was 
limited to $\lessapprox$6800 \AA. Internal quartz lamp exposures were used for flat fielding and ThAr lamp 
spectra were obtained for wavelength calibration. Integration times were 1200 s for the 2003, 2004, 
and 2005 observations and 360 s for the 2019 spectrum. The cross-dispersed spectra were reduced and 
extracted using the MAunaKea Echelle Extraction (MAKEE) pipeline written by T. Barlow.

The Near Infrared Spectrometer (NIRSPEC) operating at the Nasmyth focus of the {\it Keck II} telescope 
was used by LAH to obtain cross-dispersed, high-dispersion 1.0-1.2 $\micron$ spectra of V347 Aur on the night 
of 2019 November 25. While NIRSPEC observes the entire Y-band at $R\approx 18,500$, the spectral order 
selected for presentation here includes the Paschen $\gamma$ and the \ion{He}{1} 10830 
lines. An ABBA nod pattern was chosen for both V347 Aur and a telluric standard that was observed 
at similar airmass. The spectra were traced, extracted, wavelength calibrated, and combined using 
the REDSPEC\footnote{written by L. Prato, S.S. Kim, \& I.S. McLean} package. 

\section{Archival Data}

Hubble Space Telescope (HST) observed V347 Aur using the Wide-Field Planetary Camera 2 (WFPC2) on 
2001 July 26 as part of proposal ID number 9160 (PI D. Padgett). V347 Aur was centered on the 
Planetary Camera (PC) chip, which has a platescale of $\sim$0\farcs046 per pixel. Two deep 500 s 
exposures were made using the broadband F814W filter, saturating the core of the star, but bringing
out details of the enveloping nebulosity. Calibrated WFPC2 images were downloaded from the Mikulski 
Archive for Space Telescopes (MAST), mosaiced together, and cleaned of cosmic rays using tools 
available in the {\it stsdas} package of the Image Reduction and Analysis Facility (IRAF).

The United Kingdom Infrared Telescope (UKIRT) Infrared Deep Sky Survey, UKIDSS, (Lawrence et al. 2007), 
Galactic Plane Survey (GPS) covered $\sim$1,868 square degrees of the northern and equatorial Galactic 
plane to galactic latitudes of $-5^{\circ}<b<5^{\circ}$ in the $J, H,$ and $K$ broadband filters as well 
as $\sim$300 square degrees in the Taurus-Auriga-Perseus star forming regions. UKIDSS was completed using 
the Widefield Camera (WFCAM) and its four Rockwell Hawaii-II detectors, spaced 12\farcm83 apart, having 
a pixel scale of 0\farcs4, yielding a single exposure field of view of 0.21 square degrees. The 5$\sigma$ 
detection limits of the survey were $\sim$19.5, 19.0 and 18.4 in the $J, H,$ and $K$ passbands, respectively. 
V347 Aur was imaged by UKIDSS on 2012 January 12 using a 10 s exposure and a four position microstep pattern 
to prevent undersampling in the best seeing conditions. The pipeline reduced data from the Cambridge Astronomy 
Survey Unit (CASU) were downloaded from the Wide Field Science Archive (WSA) maintained by the Royal Observatory 
in Edinburgh, Scotland.

Low-cadence imaging of V347 Aur was obtained using the Palomar Transient Factory (PTF, Law et al. 2009) and its 7.8 square degree field of view mosaic imager coupled with the Samuel Oschin 48-inch Schmidt telescope. Images were obtained in $R-$band as well as the narrow-band H$\alpha$656 and H$\alpha$663 filters, typically two frames per night spaced by several hours. The standard integration time was 60 s in all filters, yielding a 5-$\sigma$ detection limit of 20.5 mag in $R-$band. The PTF images were downloaded from the Infrared Science Archive (IRSA) maintained at the NASA Infrared Processing and Analysis Center (IPAC) in Pasadena, California.

Regular-cadence imaging of V347 Aur was obtained from the Zwicky Transient Facility (ZTF, Bellm et al. 2019; Graham et al. 2019) archive (Masci et al. 2018)  available through IPAC. ZTF uses a custom built, wide-field charged coupled device (CCD) camera installed on the 48-inch Samuel Oschin Schmidt telescope at Palomar Observatory. The camera is composed of 16 6K$\times$6K E2V CCDs yielding an illuminated field of view of 47.7 square degrees. Three filters are available: ZTF-$g$, ZTF-$r$ and ZTF-$i$ that are designed to provide high efficiency while avoiding prominent background emission features. Exposure times for science images were 30 s, achieving a median photometric depth of ZTF-$r \sim$20.6 mag. The photometric system is calibrated to PanSTARRS, and thus in AB magnitude units.

Other archival, optical photometry was obtained from the NSVS and the ASASSN online servers. The NSVS (Wozniak et al. 2004) was conducted as part of the Robotic Optical Transient Search Experiment (ROTSE) from Los Alamos, New Mexico. ROTSE used four, co-mounted and unfiltered Cannon camera lenses, each with an 8$^{\circ}\times8^{\circ}$ field of view, imaging upon a 2K$\times$2K CCD. The platescale of each CCD was 14\farcs4 per pixel, limiting the spatial resolution of the system as well as the accuracy of the photometric calibration. Final photometry was tied to the Johnson $V-$band magnitude scale using {\it Tycho} photometry and $B-V$ colors. 

The ASASSN project (Kochanek et al. 2017) consists of multiple stations world-wide, each of which maintains a 14 cm aperture Nikon telephoto lens and a thermoelectrically cooled CCD camera. The field of view of each camera system is $\sim$4.5 deg$^{2}$, with a pixel scale of 8\farcs0 per pixel. Observations are made in the $V$ or $g$ filters, dependent upon station, and consist of three dithered 90 s exposures. 

We also examined NEOWISE data in the mid-infrared, available from IRSA (Mainzer et al. 2014).

\section{Analysis}

\subsection{Evidence for the Isolation of V347 Aur}

\begin{figure}
\includegraphics[angle=90,origin=c, height=9cm]{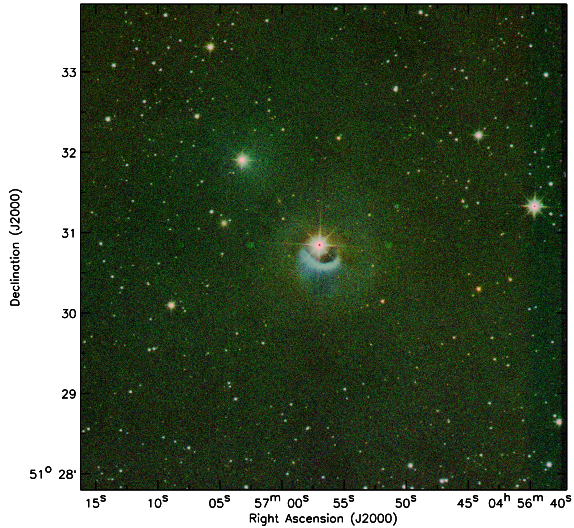}
\caption{Composite $J, H, K-$band image of V347 Aur from the UKIDSS GPS. The field of view is approximately 6\arcmin$\times$6\arcmin oriented such that north is up and east to the left. The reflection nebula enveloping V347 Aur is most prominent at $J-$band and appears to be the illuminated edge of a cavity blown out by the young stellar object.}
\label{fig:jhkimage}
\end{figure}

Given the appearance of Figure~\ref{fig:psimage} and the significant extinction suffered by V347 Aur, 
it is reasonable to assume that other pre-main sequence objects could be embedded within the L1438 molecular cloud, but not detected at optical passbands. Shown in Figure~\ref{fig:jhkimage} is a 6$\arcmin\times6\arcmin$, composite $J, H, K-$band image of L1438 obtained from the UKIDSS GPS. The small reflection nebula illuminated by V347 Aur peaks in brightness at $J-$band, appearing blue in the composite frame. The deep, near infrared image gives the impression that a toroid of molecular material envelopes V347 Aur, with its center blown out. The southern rim of the reflection nebula is illuminated by the young stellar object while its northern edge remains dark, possibly in the foreground. Several faint, reddened sources are evident in the immediate vicinity of V347 Aur, either embedded sources, or possibly, background stars and galaxies visible through the cloud.

\begin{figure}
\epsscale{1.25}
\plotone{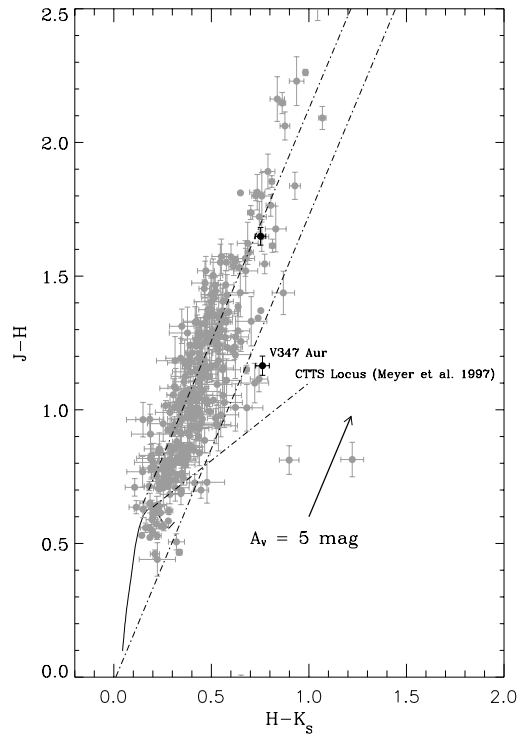}
\caption{The UKIDSS GPS $J-H$, $H-K_{S}$, color-color diagram of sources within a five arcminute radius of V347 Aur and having photometric uncertainties of $<$0.1 mag in both colors. The photometry has been placed in the 2MASS system using the transformations of Hewett et al. (2006). The dot-dashed lines represent the reddening boundaries for normal main sequence and giant stars. Also shown is the dereddened locus of Classical T Tauri stars from Meyer et al. (1997). The solid line shows the intrinsic colors of pre-main sequence (5--30 Myr old) stars taken from Pecaut \& Mamajek (2013). V347 Aur and the bright source to its northeast are both saturated in the UKIDSS data, but are plotted using 2MASS photometry and shown as black symbols.} 
\label{fig:jhhk}
\end{figure}

To distinguish candidate pre-main sequence objects from background sources, the $J-H$, $H-K_{S}$ color-color diagram of the UKIDSS GPS data is shown in Figure~\ref{fig:jhhk}. Of more than 1,000 5$\sigma$ detections within a five arcminute radius of V347 Aur, 320 have photometric uncertainties of $<$0.1 mag in both colors and are plotted in the figure. About a half dozen sources, most with significant photometric uncertainties, lie to the right of the reddening boundary for normal main sequence stars, implying near infrared excess consistent with circumstellar disk emission. These sources could represent very low-mass objects forming with V347 Aur in L1438 or, alternatively, background galaxies. Of particular interest, the bright source lying northeast of V347 Aur in Figure~\ref{fig:jhkimage}, near $\alpha$=04$^{h}$57$^{m}$03.06$^{s}$ and $\delta$=51$^{\circ}$31\arcmin54$\farcs$2 has near infrared colors consistent with a heavily reddened background giant. V347 Aur appears to be the most massive object forming within this small molecular cloud core.

Investigation of Gaia DR2 data (Gaia Collaboration 2018) confirms that, among optically visible sources, $G \lesssim 20.8^m$, V347 Aur is both spatially and kinematically isolated. Within 0.5 pc, there is only one faint ($G=20.2$ mag) source with similar parallax, but large parallax error and non-common proper motion. There are four sources with proper motions similar to V347 Aur, but all have somewhat larger proper motion values in declination, and those with measured parallax all have much larger distances than V347 Aur.

\subsection{Brightness Fluctuations of the Reflection Nebula}

The small, variable reflection nebula that is readily apparent in Figure~\ref{fig:psimage}  appears to be an illuminated wall of a cavity carved out of the enveloping molecular cloud, possibly by an outflow or jet emerging from V347 Aur. The star appears to lie in the densest region of the molecular cloud where no background stars are apparent in the Pan-STARRS $riy-$band image.

\begin{figure}
\epsscale{2.0}
\plottwo{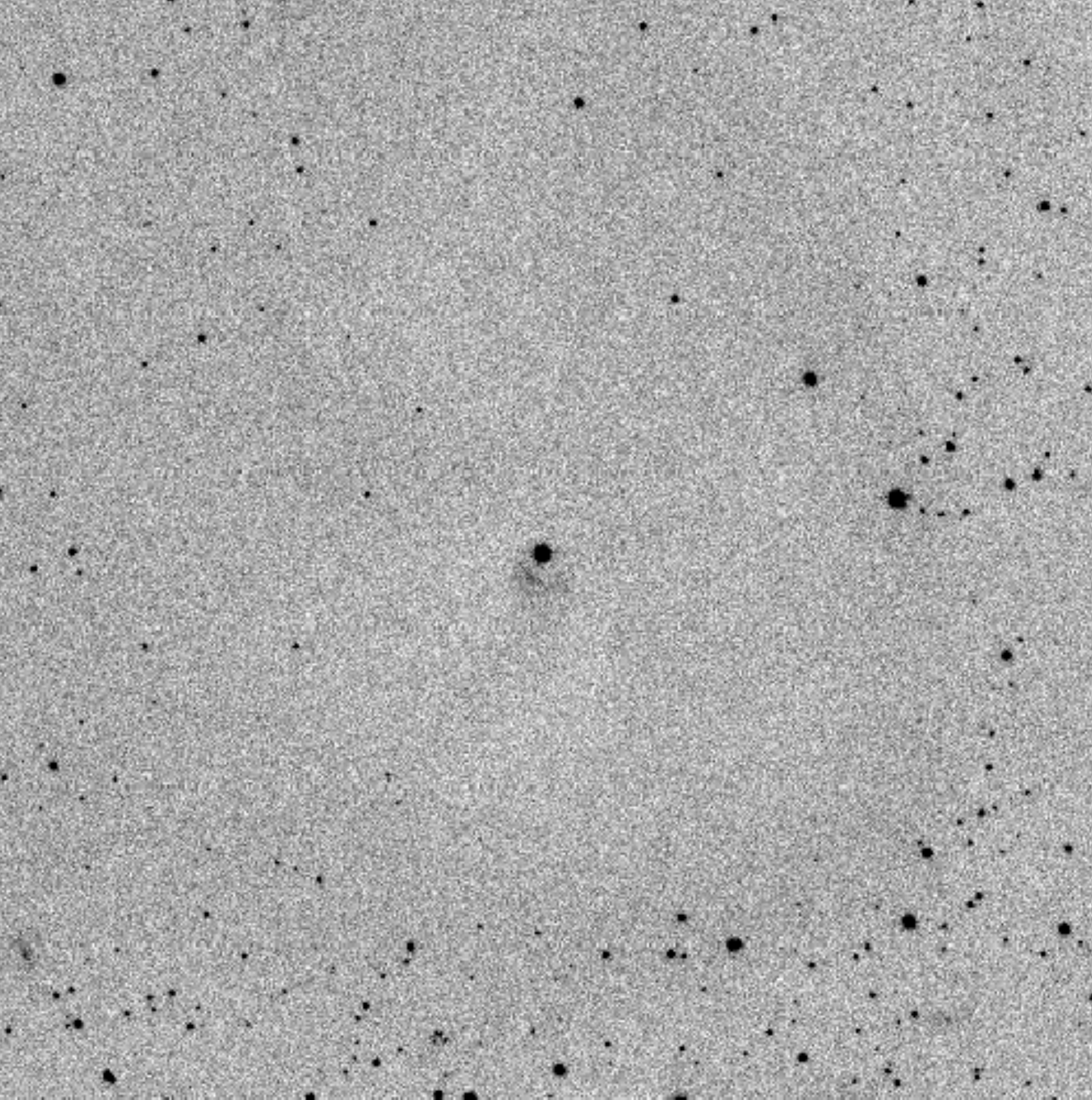}{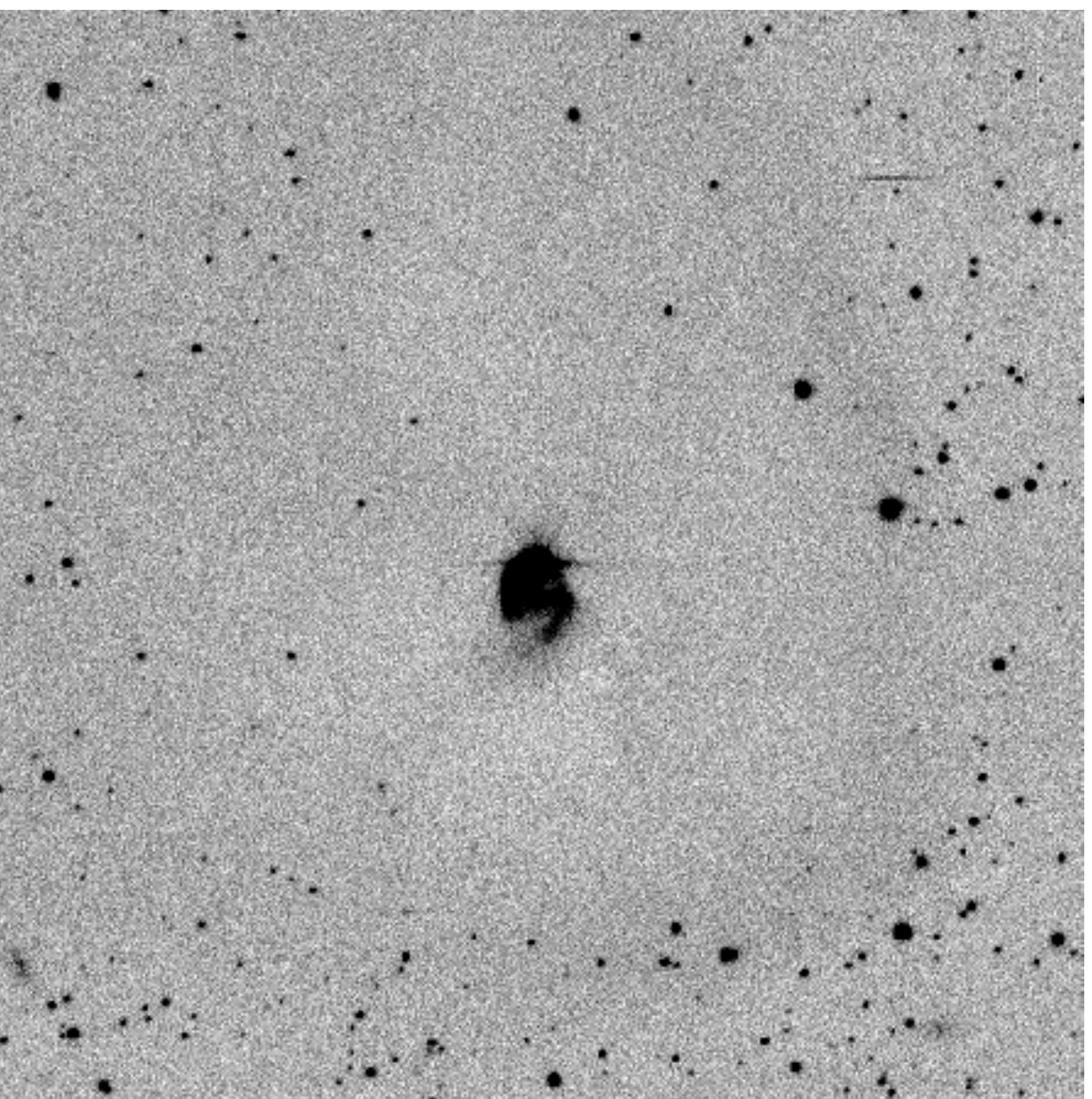}
\caption{Zwicky Transient Facility 500$\arcsec\times$500$\arcsec$, ZTF $r-$band images (north is up, east is to the left) of V347 Aur and its associated reflection nebula obtained near minimum light on 2018 September 22 (top) and after reaching peak brightness on 2018 November 10 (bottom). The brightness fluctuations of the reflection nebula are clearly correlated with those of the pre-main sequence point source.
}
\label{fig:ztfimage}
\end{figure}

The Zwicky Transient Facility imaged V347 in the ZTF-$g$ and ZTF-$r$ filters dozens of times during 2018 and 2019, capturing its rise from minimum light in 2018 September to peak brightness in 2018 November. Inspection of the broadband ZTF images confirms the brightness fluctuations in the nebula originally reported by Cohen (1978). Shown in Figure~\ref{fig:ztfimage} are two 500$\arcsec\times$500$\arcsec$, ZTF-$r$ band images centered near V347 Aur taken on 2018 September 22 and 2018 November 10. The $\sim$90$^{\circ}$ arc of nebulosity is faintly visible in the early September image when V347 Aur was near minimum brightness. This arc brightens significantly, particularly to the southeast as the pre-main sequence star reaches its maximum near ZTF-$r\sim$11.2 mag in early 2018 November. The illuminated ridge of nebulosity extends back to the south from its westernmost edge more than 30\arcsec, forming a spur in the arcuate nebula. This brightening event was related to a major maximum of V347 Aur, which was more than a magnitude brighter than previous maxima.

\subsection{Photometric Variability of the Stellar Source V347 Aur}

Multi-epoch, archival photometry of V347 Aur from ZTF, ASASSN, NSVS and the literature have revealed periodic brightness variations with amplitudes of $V\sim2.0$ magnitudes that have persisted for decades. 

\begin{figure}
\epsscale{1.25}
\plotone{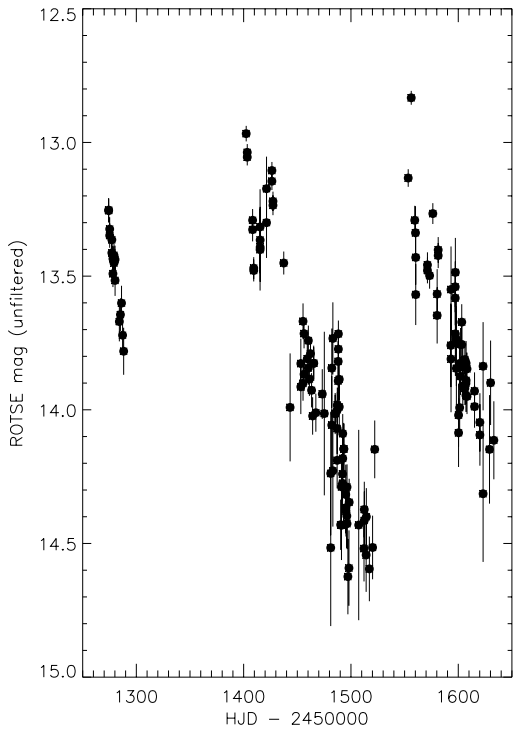}
\caption{The light curve of V347 Aur obtained from the NSVS archive 
and spanning a period from 1999 April to 2000 March. The amplitude 
and period of variation were observed to be $V\sim$1.8 mag and 150 
days, respectively. Based on these data, Wozniak et al. (2004) 
identified V347 Aur as a Mira-type variable.}
\label{fig:nsvs}
\end{figure}

Historically, the period of variability had been overestimated by a factor of two, due in part to sparse sampling or to strong period aliasing. The more recent Wozniak et al. (2004) study identified V347 Aur as a Mira-type variable and carbon star based upon its NSVS light curve, which is re-produced in Figure~\ref{fig:nsvs}. The light curve, obtained between 1999 April and 2000 March, exhibits an amplitude of $V\sim$1.8 magnitudes, varying between $\sim$12.8--14.6 mag. These changes in brightness are regular in appearance and are suggestive of a period of $\sim$150 days, close to the currently accepted period.

ASASSN has observed the V347 Aur region for the past six years, obtaining over 630 observations between 2012 February and 2018 December. The variability of V347 Aur was recognized by the ASASSN software and cataloged as a known variable. The archived ASASSN period, however, is listed as 312.85 days, which is clearly an alias.

\begin{figure}
\epsscale{1.25}
\plotone{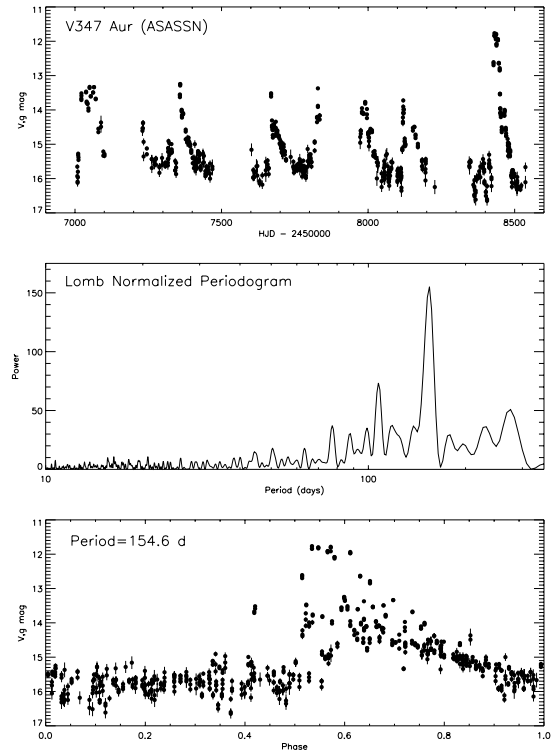}
\caption{The light curve of V347 Aur obtained from the ASASSN archive with 
photometric errors plotted for reference (top panel), the Lomb-Scargle 
periodogram (middle panel), and the phased light curve (bottom panel) for the 
peak period of 154.6 days from the periodogram. While there is a periodicity 
to the brightenings of V347 Aur, the amplitude varies from cycle to cycle.
\label{fig:asas}
}
\end{figure}

Shown in the top panel of Figure~\ref{fig:asas}  is the ASASSN $V-$band light curve of V347 Aur. Most striking is the shape of the light curve, which exhibits a steep rise in brightness of $\sim$2.5 mag occurring over $\sim$25 days, followed by a more gradual $\sim$60 day decline to minimum light. The light curve is repeated over several cycles during the four years of near-continuous observation presented in the figure. 

Shown in the middle panel of Figure~\ref{fig:asas}  is the Lomb normalized periodogram of the ASASSN photometry, which finds a strong, highly significant 154.6 day peak, consistent with the 150 day period of Wozniak et al. (2004). The phase-magnitude diagram of the ASASSN photometry for this period is shown in the bottom panel of Figure~\ref{fig:asas}. 

In late 2018 November, the peak brightness attained by V347 Aur was $\sim$1.5 magnitudes brighter than earlier maxima, reaching V$\sim$11.7 mag. This event could be analogous to a major maximum or flare-up observed periodically in EX Lupi, the prototype of eruptive pre-main sequence variables known as EXors (Herbig 2008). EXor outbursts are generally attributed to intermittent, magnetospheric accretion activity driven by circumstellar material impacting onto the stellar photosphere. The critical difference with V347 Aur, however, is that EXor outbursts are stochastic in nature, occurring at irregular intervals often separated by years. 

\begin{figure}
\includegraphics[angle=90,origin=c, height=8cm]{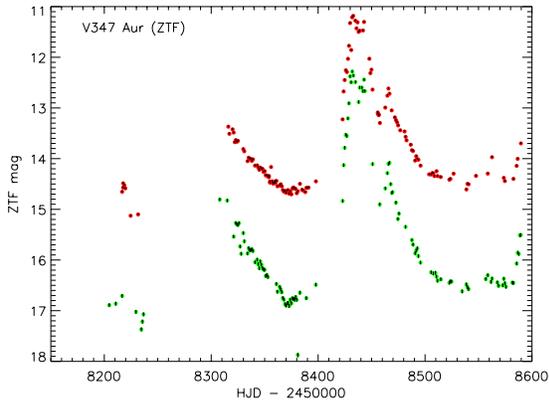}
\caption{The ZTF $g-$ and $r-$band light curves of V347 Aur obtained from the ZTF second data release. Photometric errors are overplotted, but typically are smaller than the symbol size. The observations presented here begin in 2018 March (Julian date 2458200) and extend to 2019 April (Julian date 2458590). The light curve follows the nearly five magnitude rise in brightness from a deep minimum to peak brightness in 2018 November. The ZTF curve also captures a secondary maximum that occurs some 30 days later before continuing its decline to minimum.}
\label{fig:ztf}
\end{figure}

ZTF began observing V347 Aur during the spring of 2018, imaging the region in both the ZTF-$g$ and ZTF-$r$ filters before the field passed behind the Sun. During 2018 July the field re-emerged in morning twilight and was observed consistently by ZTF on a near nightly basis, often in both filters. Shown in Figure~\ref{fig:ztf}  are the ZTF light curves of V347 Aur, which follow the pre-main sequence star's decline in brightness to an apparent minimum in 2018 September before it began brightening again. 

The steep rise in brightness and peak amplitude captured by the ASASSN photometry is confirmed by the ZTF observations. After reaching maximum in 2018 November, V347 Aur faded sharply over two magnitudes in both passbands. This decline, however, is interrupted briefly $\sim$30 days later as V347 Aur reaches a secondary maximum that is more analogous to earlier peaks in brightness. The ZTF light curve appears to experience a prolonged minimum relative to the one prior before the pre-main sequence star once again starts rising in brightness near Julian date 2458600 (2019 April).

We use the time series period finder available through the IRSA website to calculate the Lomb-Scargle periodograms for the ZTF $g-$ and $r-$band light curves. Maximum power is achieved by periods of 165.34 and 164.59 days, respectively, $\sim$10 days longer than the period identified using the significantly longer baseline ASASSN photometry. 

The ZTF photometry reveals that V347 Aur changed not only in brightness but in color as it peaked during the 2018 cycle, from ZTF-$g$=16.92 mag and ZTF $g-r$=2.26 on 2018 September 08 (minimum) to ZTF-$g$=12.37 mag and ZTF $g-r$=1.18 mag on 2018 November 10 (maximum). The two filter imaging is near simultaneous, typically several hours apart, but at times can be separated by days. V347 Aur is clearly becoming bluer as it brightens, implying a decrease in the line-of-sight extinction, the presence of a hotter source (e.g. an accretion shock or hot spot), or the occurrence of an EXor-type eruption.

The available NEOWISE photometry in the mid-infrared for V347 Aur are in the saturated regime, which is not unexpected given the Class I SED of the source, and its proximity.  Nevertheless, we find an oscillatory photometric pattern. The behavior is under-sampled on the $\sim$160 day timescale of the optical period, given the 150$-$200 day observation cadence of NEOWISE. The demonstrated oscillation is clearly an alias of the true period. A second finding from the NEOWISE data is that the mid-infrared variations appear to be colorless, in contrast to the color-magnitude trend in the optical data from ZTF. While NEOWISE data obtained near maximum are saturated as characterized by large photometric uncertainties, the W1 and W2 light curves from peak to nadir imply colorless variation with a formal color-magnitude slope of 4$\pm$8\%, i.e. consistent with zero, compared to an expected 22\% slope if due to reddening. This would support the conjecture made above that the brightness variations of V347 Aur are driven by accretion variations, more so than extinction variations.

\subsection{Stellar Properties of V347 Aur}

\begin{figure}
\epsscale{1.25}
\plotone{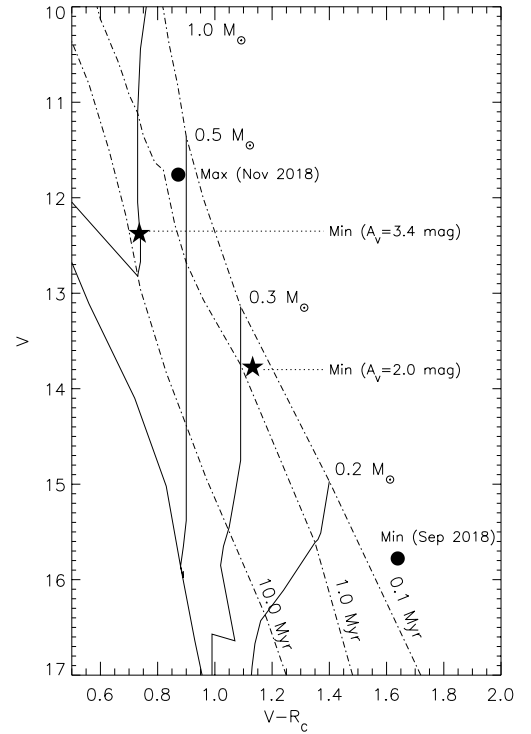}
\caption{The $V-R_{C}$, $V$ color-magnitude diagram of V347 Aur at minimum and maximum brightness plotted using photometry taken from the ZTF archive. The transformations of Tonry et al. (2012) were used to place the Pan-STARRS photometry on the Johnson-Cousins system. The pre-main sequence models and isochrones of Siess et al. (2000) are superposed. Also plotted are the extinction corrected points for photometric minimum assuming $A_{V}$=3.4 and 2.0 mag.}
\label{fig:cmd}
\end{figure}

ZTF photometry is calibrated to the Pan-STARRS photometric system, which can be placed in the Johnson-Cousins system using the transformations of Tonry et al. (2012). Shown in Figure~\ref{fig:cmd}  is the color-magnitude diagram of V347 Aur at both minimum and maximum light. Superposed are the theoretical evolutionary models and isochrones of Siess et al. (2000), which can be used to infer the physical properties of the pre-main sequence star once extinction corrections are applied.

Assuming $A_{V}\sim$3.4 mag from Connelley \& Greene (2010), the standard ratio of total to selective absorption, $R=$3.08, and adopting the {\it Gaia} parallax distance for V347 Aur of 208.96$\pm$3.95 pc, we correct the ZTF photometry for extinction at minimum light and re-plot the source in Figure~\ref{fig:cmd}. The Siess et al. (2000) pre-main sequence models predict an effective temperature of $\sim$4200 K, corresponding to a K5 spectral type. This is clearly discrepant with the M0-M2.5 spectral type inferred from the HIRES spectroscopy presented here (\S 5.5) and in the literature. The extinction suffered must be reduced significantly, to $A_{V}\sim$2.0 mag, in order for the Siess et al. (2000) models to predict an effective temperature that is consistent with the assigned spectral type. The mass and age of the resulting model are $\sim$0.35 M$_{\odot}$ and $\sim$0.9 Myr, respectively, in agreement with the models of Baraffe et al. (2015) and Feiden (2016), which were adopted by Flores et al. (2019) in their analysis.

Assuming two magnitudes of visual extinction and the {\it Gaia} parallax distance of V347 Aur, we find that in quiescence, the $V-$band magnitude, 15.77 mag, corresponds to an absolute magnitude of $M_{V}$=7.17 mag. At peak brightness during its super-maximum event in 2018 November, the absolute magnitude of V347 Aur decreased to $M_{V}$=3.16 mag, implying a change in luminosity of $\sim$4.3 L$_{\odot}$ or 1.68$\times$10$^{34}$ ergs s$^{-1}$. This change in brightness is assumed to have been induced by an accretion event. Adopting the mass (0.35 M$_{\odot}$) and radius (2.27 R$_{\odot}$) estimates of V347 Aur in quiescence taken from the pre-main sequence models of Siess et al. (2000) and using the accretion luminosity relationship of Gullbring et al. (1998), we derive an increase in the mass accretion rate in the high state of 1.1$\times$10$^{-6}$ M$_{\odot}$ yr$^{-1}$. Such a mass accretion rate would be consistent with that of an EXor during outburst (e.g. Lorenzetti et al. 2012).

\subsection{{\it Keck} High Dispersion Spectroscopy of V347 Aur}

The spectrograms of V347 Aur obtained using HIRES on {\it Keck I} on 2003 December 13, 2004 September 24, 2004 November 21, 2005 November 23 and 2019 November 29 are referred to here for convenience as epochs A, B, C, D and E, respectively. There were significant changes in V347 Aur between observations giving the distinct impression that two independent sources were present to varying degrees during all five epochs, an early M-type, absorption line spectrum and a heavily veiled, emission-line spectrum dominated by Balmer and \ion{He}{1} emission as well as optically thin \ion{Fe}{1} emission.  

\begin{figure}
\epsscale{1.25}
\plotone{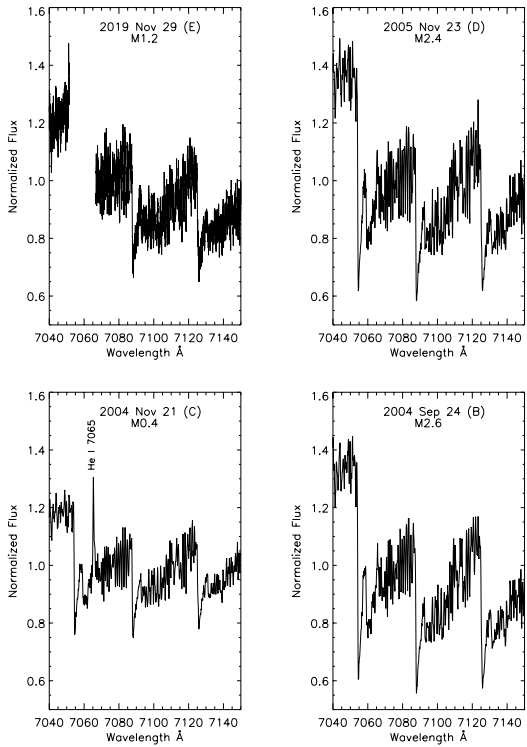}
\caption{The TiO $\lambda$7050 bandhead region for four HIRES observations of V347 Aur. Using the TiO5 index of Reid et al. (1995), which measures the depth of the bandhead relative to a defined continuum level, we find the spectral type of V347 Aur varies at least two spectral classes over months-long timescales.}
\label{fig:tio}
\end{figure}

Shown in Figure~\ref{fig:tio}  are the TiO $\lambda$7050 bandhead regions for four of the HIRES spectra (the 2003 spectrum, A,  does not extend to this wavelength). The spectra reveal significant variation in the depths of the continuum reached by the TiO bandhead. The TiO5 spectral index of Reid et al. (1995) determines the ratio of fluxes between the continuum measured near $\lambda$7044\AA\ and the depth of the strongest TiO feature at $\lambda$7130\AA. The index is well calibrated with spectral type and is found to be effective between types $\sim$K7 and M6.

Adopting the TiO5 index and its linear relationship with spectral type (Reid et al. 1995), we estimate the spectral type for each available HIRES spectrum to be: epoch B: M2.6, C: M0.4, D: M2.4, and E: M1.2. The uncertainties associated with the TiO5 index are approximately half a spectral class. Even the sparsely sampled HIRES observations demonstrate changes in spectral class occurring over timescales as short as two months. 

Cross correlations between the spectrum from epoch E and a library of radial velocity standards taken with the same spectral setup of HIRES, lead to an inferred heliocentric radial velocity for V347 Aur of $+8.9\pm 0.8$ km s$^{-1}$.

Classic signatures of accretion activity include strong H$\alpha$ emission (Bertout 1989), low excitation, forbidden line emission (Edwards et al. 1987; Simon et al. 2016), as well as \ion{He}{1} and \ion{Ca}{2} emission. The filling in of photospheric absorption lines, or veiling, is attributed to an optical and UV continuum excess emission arising from an accretion hot spot, where material is impacting onto the stellar photosphere (e.g. Hartigan et al. 1991). 

\begin{figure}
\epsscale{1.25}
\includegraphics[angle=0,origin=c, width=\columnwidth]{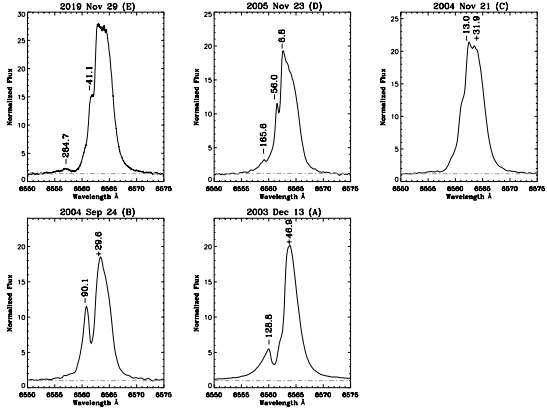}
\caption{{\it Keck} HIRES spectra of V347 Aur centered near H$\alpha$. Velocities of prominent features (in km s$^{-1}$) in the emission line profiles are annotated as is the approximate continuum level (dot-dashed line). The strength of the H$\alpha$ emission line in all five epochs is indicative of magnetospheric accretion. Of note are the blue-shifted absorption features present in spectra A, B, D, and E, which are suggestive of strong disk winds.
}
\label{fig:halpha}
\end{figure}

The {\it Keck} high dispersion spectra of V347 Aur reveal strong H$\alpha$ emission in all five epochs of observation, ranging in strength from FWHM$\sim$65\AA (B) to 86\AA (C). The H$\alpha$ profiles from the five epochs of observation are shown in Figure~\ref{fig:halpha} , which also annotates the heliocentric radial velocities of various features in the emission line structure. The strong H$\alpha$ emission serves as a probe of infalling gas from the accretion disk onto the pre-main sequence star. From the strength of emission and the structure of the H$\alpha$ profiles, we can infer that magnetospheric accretion is occurring during all five epochs.  

A blue-shifted absorption component is evident in several of the H$\alpha$ profiles shown in Figure~\ref{fig:halpha}. This feature is generally attributed to strong disk winds, but its appearance in some but not all of the spectra could imply a variable mass loss rate from the disk. Table 1 provides the measured equivalent widths and radial velocities of specific absorption and emission features for each of the five epochs of high dispersion spectroscopy including H$\alpha$, \ion{He}{1} $\lambda\lambda$5876, 6678, \ion{Na}{1} D lines, [O I] $\lambda\lambda$6300, 6363, and [S II] $\lambda\lambda$6717, 6731. The radial velocities given are approximately heliocentric, but were determined purely from the wavelength calibration procedure and the applied barycentric corrections. The associated error with the radial velocities is expected to be 1--2 km s$^{-1}$.

The \ion{Na}{1} D lines ($\lambda\lambda$5890, 5896) are found in emission and significantly broadened (FWHM $\sim$100 km s$^{-1}$) in all five epochs of HIRES spectra. The narrow interstellar absorption components have heliocentric radial velocities of $\sim+$5 km s$^{-1}$, consistent with the velocity of the L1438 molecular cloud. 

\begin{figure}
\epsscale{1.25}
\plotone{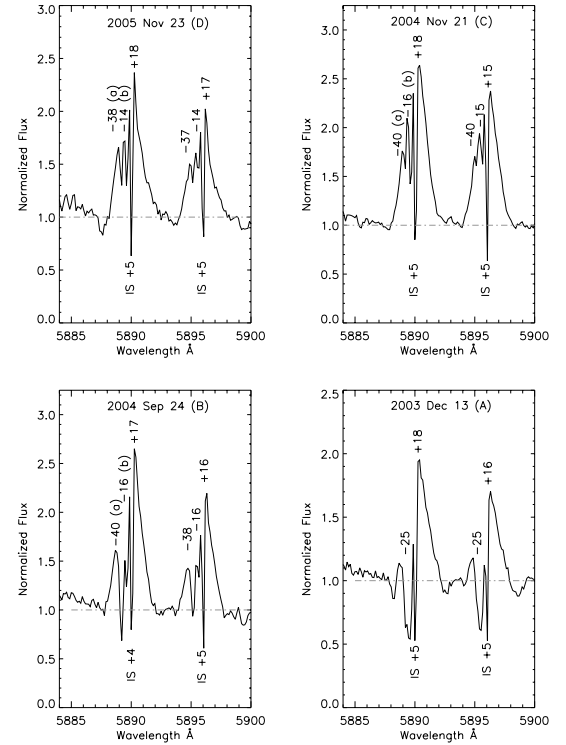}
\caption{The Na I D line spectral region of V347 Aur for four epochs of observation. The narrow interstellar lines are superimposed upon the broad emission line profile of the star with radial velocities of $\sim$+5 km s$^{-1}$, consistent with the radial velocity of the L1438 molecular cloud. Also present are components of an outflowing disk wind, labeled as `a' and `b' in each panel around $\lambda$5889 (with the exception of spectra A and E where only one component is resolved). Heliocentric radial velocities of prominent features (in km s$^{-1}$) are indicated.}
\label{fig:nad}
\end{figure}

Pascucci et al. (2015) suggest that magnetospheric accretion is responsible for the broad \ion{Na}{1} D emission line profiles in T Tauri stars having high mass accretion rates. Similar to H$\alpha$ emission, the \ion{Na}{1} D line emission traces the in-falling gas accreting from the circumstellar disk onto the stellar photosphere. Within the \ion{Na}{1} D emission lines in V347 Aur are two narrow absorption components at approximately $-$39 and $-$16 km s$^{-1}$ that vary in strength and structure. These absorption line profiles may be linked with outflows or disk winds. Shown in Figure~\ref{fig:nad}  are the \ion{Na}{1} D line regions of four of the HIRES spectra with heliocentric radial velocities of prominent features annotated. Spectrum E (2019 Nov 29) is not presented here given that the \ion{Na}{1} D lines are split between two spectral orders, but its appearance is similar to the other epochs.

In addition to the lines highlighted above, V347 Aur also exhibits broad emission from a plethora of \ion{Fe}{1} and \ion{Fe}{2} lines throughout the optical spectrum, as well as from the \ion{Mg}{1}b triplet and \ion{O}1 $\lambda 8446$.

Forbidden line emission in classical T Tauri stars is generally attributed to outflows or disk winds, which remove excess angular momentum resulting from material accreting onto the photosphere of the pre-main sequence star (Pascucci et al. 2015). The HIRES spectra of V347 Aur all exhibit strong [S II] $\lambda\lambda$6717, 6730 and [O I] $\lambda\lambda$6300, 6363 emission as well as weaker [N II] $\lambda\lambda$6548, 6583 and [Fe II] $\lambda$7155 
emission. These transitions are optically thin, and are generally composed of a high velocity component (HVC) associated with collimated jets and a low velocity component (LVC) that originates from a disk wind.

The equivalent widths of the [S II] and [O I] emission lines in V347 Aur are greatest when the M-dwarf photosphere is most prominent (i.e. quiescence). Both sets of forbidden lines are doubled, the LVCs of [S II] having heliocentric radial velocities of $-$18 km s$^{-1}$ and half the strength of the HVC at $-$90 km s$^{-1}$. The LVCs of [O I] $\lambda$6300, 6348 have heliocentric radial velocities of $-$5 km s$^{-1}$ and are comparable in strength to their HVCs at $-$95 km s$^{-1}$. Neither [O I] nor [S II] emission lines extend a detectable distance from the continuous spectrum on any HIRES spectrogram; an offset at right angles to the dispersion direction of as little as 0\farcs3 would be apparent (Herbig, private communication). Only the HVC of [N II] $\lambda$6583 is measurable, having a radial velocity of $-$92 km s$^{-1}$.

\begin{figure}
\epsscale{1.25}
\plotone{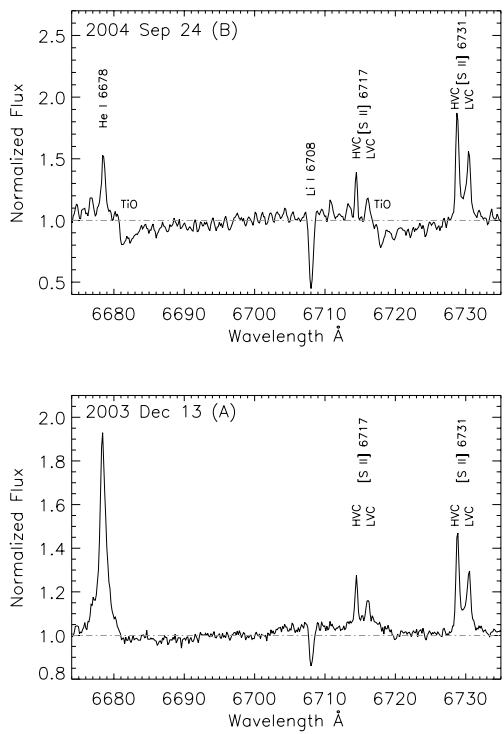}
\caption{The spectral region near [S II] $\lambda\lambda$6717, 6731 for epoch A (bright state) and epoch B (quiescent state). Included in this wavelength range is the He I 6678 emission feature and the Li I 6708 absorption line. HVC and LVC components for [S II] $\lambda\lambda$6717, 6731 are identified. Note the filling-in of the TiO bandheads and the Li I 6708 absorption line during outburst.}
\label{fig:helisii} 
\end{figure}

Shown in Figure~\ref{fig:helisii}  are the wavelength regions near the [S II] $\lambda\lambda$6717, 6731 emission features for spectrum A (bright emission state) and spectrum B (quiescent state). These regions include the \ion{He}{1} $\lambda$6678 emission line and the \ion{Li}{1} $\lambda$6708 photospheric absorption line. The TiO bandheads centered near 6680 and 6714 \AA\ are readily apparent in the quiescent spectrum, but appear filled-in in spectrum A. Similarly, the depth and equivalent width of the \ion{Li}{1} $\lambda$6708 absorption feature is considerably reduced in the bright emission spectrum, presumably from continuum excess emission. In the "quiescent" spectra B and D when the M-type absorption spectrum dominates, EW(Li I)$\sim$0.4 \AA. In the emission dominated spectrum A, however, EW(Li I)$\sim$0.1 \AA\ is measured, while 0.3 \AA\ is found for the intermediate states of the C and E spectra (see Table 1). 

 The \ion{He}{1} $\lambda$6678 emission line appears significantly stronger in spectrum A than in B and possesses a broad component that is over two hundred km s$^{-1}$ in width, possibly arising from the accretion column. The central core of the feature, however, is consistent with the heliocentric radial velocity of the pre-main sequence star, suggesting that its origin is near the stellar surface, perhaps the accretion shock.

\begin{figure}
\includegraphics[angle=-90,origin=c, width=\columnwidth]{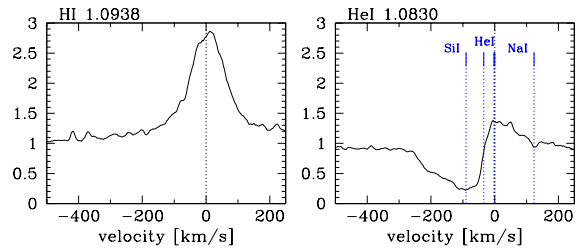}
\caption{Line profiles from a single order of the high dispersion {\it Keck II} NIRSPEC spectrum of V347 Aur. 
Shown are the Pa $\gamma$ (left panel) and the \ion{He}{1} $\lambda$10830 (right panel) profiles. Vertical dotted lines mark the feature rest wavelengths. Note that the \ion{He}{1} line is a triplet, with two closely spaced features at and near zero, and the third weaker component slightly blueshifted.  Although the profile is dominated by the \ion{He}{1} contribution, there is possible minor absorption from nearby \ion{Na}{1} and \ion{Si}{1} features as well.}
\label{fig:nirspec}
\end{figure}

NIRSPEC on {\it Keck II} was used to probe the \ion{He}{1} $\lambda$10830 emission line at high dispersion, which is sensitive to both the accretion column and disk winds. Shown in Figure~\ref{fig:nirspec} is one order of the NIRSPEC spectrum of V347 Aur centered near \ion{He}{1} $\lambda$10830, which includes Paschen $\gamma$. The P Cygni-like profile of the \ion{He}{1} $\lambda$10830 feature shows deep, blue-shifted absorption that extends well below the continuum and out to radial velocities of $-$250 km s$^{-1}$ from line center. The \ion{He}{1} $\lambda$10830 profile is clearly distinct from those of \ion{He}{1} $\lambda$5876 and $\lambda$6678 (see Figure~\ref{fig:helisii}), which exhibit broad, single peaked emission in the HIRES spectrum obtained four nights later. The origin of the \ion{He}{1} $\lambda$10830 blue-shifted absorption is likely strong stellar as well as disk winds (e.g. Edwards et al. 2006).

We note that both the NIRSPEC and the HIRES spectra from late 2019 were taken during a brightness maximum, according to available photometry.  This local peak was much smaller than the major eruption that occurred in late 2018, however. 

The variations observed in the spectra of V347 Aur between quiescence and its outburst state are not unique. Giannini et al. (2017) report similar variations in the classical EXor V1118 Ori, which entered an outburst state in 2015. The mass accretion rate of V1118 Ori was observed to increase by two orders of magnitude, during which its optical spectrum exhibited an abundance of metallic emission lines. Other examples of pre-main sequence objects having spectra that varied between quiescence and outburst include PTF 10nvg (Hillenbrand et al. 2013), PTF 15afq (Hillenbrand 2019), and ASAS 13db (Sicilia-Aguilar et al. 2017).

\subsection{Constraints on Multiplicity}

\begin{figure}
\includegraphics[angle=0,origin=c, height=12cm]{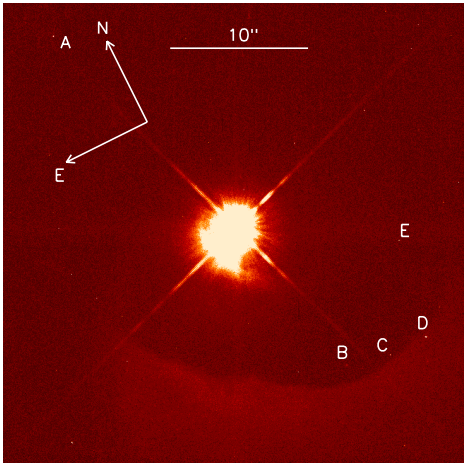}
\caption{A subsection of the deep WFPC2 F814W image of the reflection nebula and V347 Aur centered on the PC chip. The inner edge of the reflection nebula is visible to the south and southeast of V347 Aur. The arcuate nebula lying close in to the pre-main sequence star is readily apparent. Several faint sources labeled A$-$E lie near the inner edge of the reflection nebula, but are likely background objects.} 
\label{fig:hstimage}
\end{figure}

One possible explanation for the periodic variability of V347 Aur is the presence of a stellar or sub-stellar companion that induces accretion instabilities as it nears periastron. The HST WFPC2 imaging of V347 Aur provides a means of probing for a spatially resolved companion of the pre-main sequence star, particularly at separations $>$0\farcs5. The combined HST WFPC2 F814W images of V347 Aur are shown in Figure~\ref{fig:hstimage}, stretched logarithmically and scaled to highlight the structural details of the nebulosity. 

The faintly luminous edge of the reflection nebula is evident to the south and southeast, approximately $\sim$10\farcs5 from the star or $\sim$2,200 AU in projected separation. An extended gap of $\sim$8\arcsec exists between the inner edge of the reflection nebula and a protuberance of nebulosity that appears to be physically associated with the star. This short arc of nebulosity, $\sim$1\farcs8 or $\sim$370 AU from V347 Aur, extends for approximately six arcseconds and runs roughly north-south. At its broadest point, the arcuate nebula is $\sim$60 AU wide (projected), and may represent remnant material from the collapse and formation of the star and its circumstellar disk.

Several faint sources labeled A$-$E in Figure~\ref{fig:hstimage}  are identified near the inner edge of the reflection nebula, but are more than 10\farcs0\ from V347 Aur. With only two deep WFPC2 frames available, cosmic ray rejection is problematic, but these five sources appear in both frames and have point-source-like point spread functions (PSF). At least one, object D, may have been detected in the deep UKIRT near infrared imaging of the region. If not artifacts, these five sources are likely background objects visible through the nebulosity, but even if associated with V347 Aur, they are much too distant to be responsible for the brightness variations observed to occur over $\sim$160 day timescales.

V347 Aur is saturated in the HST WFPC2 frames, making a companion search within the inner $\sim$0.5\arcsec ($\sim$100 AU) problematic. The wings of the PSF, however, do not appear elongated or enhanced in any preferential direction as might be expected if a bright companion were present. The arcuate nebula raises the background level near the star, further complicating the search for close, fainter companions. 

Connelley et al. (2009) used the Subaru adaptive optics (AO) system with the coronagraphic imager to examine V347 Aur at high angular resolution in $L'$-band. The thermal AO imaging should have been capable of resolving a close ($<$ 100 AU) companion to the pre-main sequence star, but the authors reported no detections.

While the question of a spatially resolved companion to V347 Aur is at least partially answered, the five epochs of {\it Keck} high dispersion spectroscopy allow for an examination of spectral lines for doubling or radial velocity variations that could be attributed to the presence of a companion. The \ion{Li}{1} $\lambda$6708 photospheric absorption feature shown in Figure~\ref{fig:helisii} is clearly narrow and exhibits no evidence of line doubling. Thorough inspection of the HIRES spectrograms and the few photospheric features available within them reveals no evidence of spectral line doubling or broadening that could be attributable to a stellar mass companion. 

Cross-correlating select orders of the HIRES spectra with each other, we find no detectable shifts among the spectra taken months or even years apart, i.e. the offsets in radial velocity between observations are consistent with zero. From this result, we conclude that V347 Aur does not experience radial velocity variations above our detection threshold of $\sim$2.0 km s$^{-1}$, which is comparable to the typical radial velocity "noise" level of young pre-main sequence stars (e.g. Hillenbrand et al 2015). It is certainly possible, however, that the orbital plane of a hypothetical companion is perpendicular to or nearly so to Earth's line of sight, thereby reducing its radial velocity signature.
 
Similarly, several orders from the high dispersion, near infrared spectrum of V347 Aur presented by Flores et al. (2019) do not reveal evidence for binarity at $K-$band, where the contrast ratio between the primary and a potential lower mass companion would be significantly reduced. The photospheric absorption lines of \ion{Ca}{1}, \ion{Al}{1} and \ion{Na}{1} appear single and minimally broadened. This is further supported by the authors' vsin($i$) measurement of 11.7 km s$^{-1}$, which is consistent with or below the mean rotational velocity of pre-main sequence M-dwarfs. 
 
\section{Discussion}

\subsection{Isolated Star Formation in the L1438 Molecular Cloud}

Inspection of the composite, Pan-STARRS $riy-$band image of V347 Aur presented in Figure~\ref{fig:psimage}  suggests that 
V347 Aur is the only star within the L1438 molecular cloud that is obviously interacting with nebular material. V347 Aur appears to lie in the densest region of the cloud, where no background stars or galaxies are visible at optical wavelengths. There are several stars that appear heavily reddened by the cloud, particularly along its periphery. In \S 5.1 we found a lack of evidence for infrared excess sources across the L1438 cloud.  V347 Aur thus appears to be the only source in the vicinity with hot circumstellar dust.  

Herbig (private communication) stated that there were also no other H$\alpha$ emission sources in L1438, implying that he had surveyed the cloud using either the wide-field grism spectrograph on the University of Hawaii 2.2 m telescope on Maunakea or the grating spectrograph on the Crossley reflector at Lick Observatory. Herbig's statement appears to be re-affirmed by differencing narrow-band H$\alpha$656 and continuum H$\alpha$663 images obtained from the PTF archive, and identifying sources with residual H$\alpha$ emission. Only V347 Aur remains in the subtracted frame, implying strong H$\alpha$ emission is present in this source alone.

Whether V347 Aur is forming in true isolation or is part of a larger association, is not well established. 
To assess the broader context, we consider that the {\it Gaia}-measured parallax and proper motion of V347 Aur are:
$\pi$=4.79$\pm$0.09 mas, corresponding to a distance of 208.96$\pm$3.95 pc, and $\mu$($\alpha$)=+0.66$\pm$0.15 mas yr$^{-1}$, $\mu$($\delta$)=$-$20.87$\pm$0.14 mas yr$^{-1}$. V347 Aur therefore lies well in the foreground of the Camelopardalis OB1 
association, which has a distance of 1010$\pm$210 pc (Straizys \& Laugalys 2007), as well as other nearby, young galactic 
clusters including NGC\,1502, NGC\,1545, NGC\,1528, and NGC\,1513.

\begin{figure}
\includegraphics[angle=90,origin=c, height=9cm]{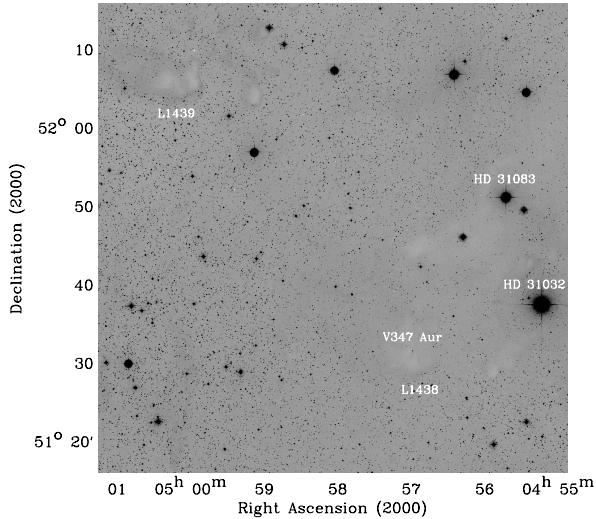}
\caption{The blue Palomar Sky Survey plate for a one square degree area around V347 Aur. To the northeast lies the Bok globule L1439. The two bright A0 V stars that could illuminate the western edges of L1438 and L1439 are labeled. V347 Aur and its parent molecular cloud core are located near bottom right.}
\label{fig:possimage}
\end{figure}

Shown in Figure~\ref{fig:possimage} is the blue Palomar Sky Survey plate for a one square degree area in the vicinity of V347 Aur. Approximately half a degree to the northeast of the pre-main sequence star lies the Bok globule L1439 or CB26 (Clemens \& Barvainis 1988). As described by Reipurth (2008), this Bok globule is about five arcminutes in diameter with a faint, luminous rim facing west. The millimeter imaging of Launhardt \& Sargent (2001) 
identified a deeply embedded, $\sim$0.3 M$_{\odot}$ protostar with an edge-on disk forming within L1439. These authors suggested the cloud may be an extension of the Taurus-Auriga star forming region (see also Sadavoy et al. 2018)

The western and southwestern edges of the L1438 core hosting V347 Aur appear, similar to L1439, faintly luminous (see Figure~\ref{fig:possimage}), as if illuminated by a nearby, bright star. The obvious candidate for the source of illumination is HD 31032 ($V$=7.04 mag, A0 V) located approximately 20$\arcmin$ west-northwest of V347 Aur. The {\it Gaia} measured parallax for HD 31032 is $\pi$=6.55$\pm$0.05 mas, corresponding to a distance of 152.66$\pm$1.25 pc, some $\sim$50 pc in the foreground of L1438. The proper motion of HD 31032, $\mu$($\alpha$)=+12.79$\pm$0.08 mas yr$^{-1}$, $\mu$($\delta$)=$-$23.71$\pm$0.06 mas yr$^{-1}$, is only marginally consistent in $\mu$($\delta$) with V347 Aur. 

Approximately fifteen arcminutes north-northeast of HD 31032 is a second A0 V star, HD 31083 ($V=$8.03 mag), with a {\it Gaia} measured parallax and proper motion of $\pi$=5.25$\pm$0.05 mas, corresponding to a distance of 190.53$\pm$1.93 pc, and $\mu$($\alpha$)=+4.84$\pm$0.07 mas yr$^{-1}$, $\mu$($\delta$)=$-$29.71$\pm$0.05 mas yr$^{-1}$. The distance of HD 31083 is more consistent with that of V347 Aur, as is its measured proper motion, though they are not formally co-moving. 

It is suggested that one or more of these early-type stars may be associated with the molecular cloud complex comprising L1438 and L1439, and could represent the most massive members of a loose young stellar association that is still in the process of formation.

We can also consider the possible relationship between V347 Aur and the Taurus-Auriga young stellar complex.  Although in a similar part of the sky,
the northern-most aggregates of the Taurus-Auriga molecular clouds are located $\sim$20$^{\circ}$ south of the L1438 and L1439 cores, implying a 
projected separation $\sim$50 pc. The parallax of V347 Aur yields 209 pc, a somewhat larger distance than found for the stellar population of Taurus-Auriga at $\sim$130$-$160 pc for the different sub-groups (Roccatagliata et al. 2020). The proper motion of V347 Aur is also formally inconsistent, though similar to the mean of Taurus-Aurgia members, which Roccatagliata et al. (2020) reports as ranging among $\mu$($\alpha$)=4.6 to 12.4 mas yr$^{-1}$ and $\mu$($\delta$)=$-$17.2 to $-$25.4 mas yr$^{-1}$ for the various sub-groups.

Whether the L1438 and L1439 molecular cloud cores and their embedded pre-main sequence stars are part of the Taurus-Auriga complex, or members of a distinct stellar association, warrants further investigation, but is beyond the scope of the current work. What can be stated is that V347 Aur is relatively isolated, a very different environment from that experienced by the vast majority of stars that form in densely populated, cluster environments.

\subsection{The Variable Nebula}

Several variable reflection nebulae are associated with pre-main sequence stars.  These include the famous Hubble's Variable Nebula (Hubble 1916), the R Coronae Australis reflection nebula (Knox Shaw 1916), and more recently, one associated with HBC 340 in NGC\,1333 (Hillenbrand et al. 2015; Dahm \& Hillenbrand 2017). The sources of variability for these reflection nebulae are generally attributed to dust casting shadows onto the nebula wall as in the case of R Mon and Hubble's Variable Nebula, or circumstellar material eclipsing the illuminating star analogous to HBC 340. 

The V347 Aur nebular variations can be seen on the IRSA ZTF image viewer, and were highlighted in Figure~\ref{fig:ztfimage}.
The brightness fluctuations of the reflection nebula seem immediately
correlated with the pre-main sequence star's variability. 
One interpretation is that we have a view to the source that is relatively 
pole-on, through the foreground molecular cloud core and dusty circumstellar 
envelope, directly toward the hot origins of the photometric variations.

\subsection{The Origin of the Variability in V347 Aur}

The variability of V347 Aur was originally classified by Wozniak et al. (2004) as long period or Mira-type, given the shape and periodicity of its light curve and the star's M-type spectral classification. The median absolute magnitude of evolved, Mira-type variables, however, is nearly two magnitudes brighter than V347 Aur at maximum. Given its association with the L1438 molecular cloud, the presence of numerous spectroscopic accretion signatures, and its strong \ion{Li}{1} $\lambda$6708 absorption line profile, V347 Aur is unambiguously a pre-main sequence object.

The origin of variability in pre-main sequence stars is generally attributed to rotating star spots, irregular variability arising from stellar flare events, accretion activity, obscuration from circumstellar material, or eclipsing companions. In the case of V347 Aur, although the light curve is periodic, the period is unusually long for a young star, and the variations have unusually large-amplitude. Spot-induced rotational modulation is thus not a viable explanation for the observed brightness variations. More stochastic processes such as flare events, accretion-induced outbursts, or extinction phenomena are better able to explain the large amplitude, but have more difficulty accounting for the regular periodicity of V347 Aur. If an eclipse were responsible, the eclipse profile would have a highly unusual depth and duration, and would need to involve extended circumstellar material around one or both objects. Furthermore, we have argued above that the evidence for binarity is not strong. 

In addition to accretion phenomena, eruptive pre-main sequence stars may also suffer from variable extinction as noted by Jurdana-{\v{S}}epi{\'c} et al. (2017). Obscuration from circumstellar material passing in front of V347 Aur, however, cannot account for the changing spectral features observed between photometric minimum and maximum. While local extinction effects may play some role in the variability of V347 Aur, they are certainly not the dominant source.

To further constrain the origin of the V347 Aur variations, we must consider the evidence from both the spectroscopic and the photometric observations. There is a veiled continuum and a bright emission-line spectrum associated with the bright photometric state. The light curve profile is another important piece of evidence, with Figure~\ref{fig:asas} demonstrating a repeating pattern of $\sim$30-day rise times and $\sim$60-day decay times. There are also color changes. The limited near-simultaneous, multi-color ($g-$ and $r-$band) photometry from the ZTF archive demonstrates that V347 Aur reddens as it fades, with bluer color evident during brightness maxima. 

The body of observations that we have presented all suggest that a hot component becomes visible as the brightness of V347 Aur increases. This is perhaps an enhanced accretion column or a hot spot where circumstellar material is impacting onto the stellar photosphere. Repeating accretion burst events could account for the sharp rises in brightness and the varying peak brightness levels from cycle to cycle. 

The extended decay times imply that the cooling timescale is approximately a factor of two longer than the heating timescale. A rapid increase in the mass accretion rate as material flows from the inner circumstellar disk onto the stellar photosphere would account for the short duration of the rise in brightness. Lee et al. (2020) suggest that the prolonged decay timescale of the similar periodic, large amplitude eruptive variable V371 Ser is associated with the viscous spreading time within its accretion disk. 

Present in the high dispersion spectra are high velocity, blue-shifted emission lines that can be attributed to a jet or a wind, suggesting that accretion is strong. The low measured vsin($i$) implies slow rotation and possibly argues for a near pole-on geometry. An open cavity into the L1438 molecular cloud aligned close to our line of sight would also explain the brightness variations in the extended nebulosity.

The periodicity of V347 Aur's photometric behavior suggests that there are further special circumstances to the general repeating accretion burst scenario outlined above. Other examples of quasi-periodic, discrete accretion bursts  were identified in a sample of Upper Scorpius pre-main sequence stars by Cody et al. (2017).  These bursts have somewhat smaller amplitudes (0.5$-$2 mag), shorter  rise-to-fall timescales ($\sim$1 week), and less extended decay profiles than V347 Aur. 

More strictly periodic ``pulsed accretion" has also been identified in several well-studied close binary systems, including DQ Tau (Mathieu et al, 1997) and TWA 3A (Tofflemire et al. 2017, 2019). As pointed out by Tofflemire et al. (2019), these are both high eccentricity binaries with $e\approx 0.6$. The lower eccentricity, $e\approx 0.3$, close binary system UZ Tau E studied by Jensen et al. (2007) and Ardila et al. (2015), does not exhibit the same coordination between the binary orbit and accretion variations.  Jensen et al. (2007) discuss several other systems with $0.4 < e < 0.5$ which lack strong evidence for impulsive accretion. The suggestion is that high eccentricity is required in order to bring the components close enough together, perhaps within only a few $R_{star}$ of one another, in order to periodically influence the disk-to-star accretion flow.  Circumbinary accretion scenarios as a function of eccentricity and mass ratio have been studied in detail recently by e.g. Munoz et al. (2020) and Duffell et al. (2020). We highlight that all of the exemplar pulsed accretion binaries have shorter periods and lower amplitude variability than the case of V347 Aur, and that the binary simulations also predict maximum accretion rate variations that do not exceed 50$-$100\%.

The large amplitude of the V347 Aur photometric behavior, factors of several to tens in brightness, suggests that the source could be experiencing periodic outbursts driven by accretion disk instabilities. The regularity of the bursts implies that they could be induced by a companion, though one located further out in the disk than in the close binary scenario discussed above.  Such an episodic accretion scenario could readily account for the changing spectrum of V347 Aur at outburst, the color differences between maximum and minimum light, and its regular periodicity. The variable mass accretion rate would be regulated in part by the binary orbit. Near periastron, the mass accretion rate would increase, resulting in photometric brightening, increased continuum veiling and enhanced line emission. 

V371 Ser (Hodapp et al. 2012; Lee et al. 2020) is perhaps in the same category as V347 Aur, although with a significantly longer period, $\sim$530 days. Lee et al. (2020) propose that circumstellar material builds up within the inner $\sim$0.4 AU of the disk before becoming unstable and flowing rapidly onto the stellar photosphere. The increase in luminosity is attributed to a nearly order of magnitude increase in mass accretion rate. The triggering mechanism for the periodic eruptions in V371 Ser, however, remains unknown.

Kuruwita et al. (2020), and previously Lodato \& Clarke (2004), considered from a theoretical perspective binary- and planet-induced episodic accretion scenarios at early stages.  These may be applicable in modified form for the case of V347 Aur. Reipurth (2000) addresses a similar mechanism in which a companion in an eccentric orbit modulates the mass accretion rate onto the primary star, leading to regulated outflow activity.

Despite the suggestive theoretical underpinnings discussed above,
the evidence presented here, however, does not support the presence of a stellar companion to V347 Aur that is either spatially or spectroscopically resolved. Speculating that the photometric period of V347 Aur is correlated with a close, undetected sub-stellar mass companion, and assuming a theoretical mass of $\sim$0.3 M$_{\odot}$ for V347 Aur, we find that the $\sim$160-day period corresponds to a Keplerian orbital radius of $\sim$0.4 AU. Using the 2 km s$^{-1}$ detection threshold from the cross-correlation analysis as a velocity semi-amplitude, and assuming the eccentricity of the hypothetical orbit to be negligible, we use equation 1 of Cumming et al. (1999) to determine the upper limit for the companion mass of V347 Aur to be $\sim$24 M$_{J}$ sin($i$), where M$_{J}$ is the mass of Jupiter and $i$ is the inclination angle of the orbit.

Other mechanisms that could be responsible for the periodic nature of V347 Aur's outburst phenomena include large-scale structures within the circumstellar disk that influence mass transport over viscous timescales (Hartmann et al. 2016). Such disk instabilities are thought to be responsible for the months-long and multi-year to multi-decade long outbursts observed in EXor and FU Ori class variables, respectively. The shorter timescales of EXor outbursts point to a build-up of material in the inner disk region that outpaces the flow of gas onto the stellar photosphere. When gas density and pressure become adequately high, material flows rapidly onto the stellar surface, inducing the dramatic rise in mass accretion rate and brightness. V347 Aur could be a regularly repeating EXor type system with somewhat sharper, rather than plateauing, light curve peaks. 

\section{Summary}

V347 Aur is an embedded, late-type (M0$-$M2) pre-main sequence star forming within a small core in the L1438 molecular cloud that experiences periodic brightness variations of amplitude $V\sim$2.0 magnitudes or greater with a $\sim$160 day period. These brightness fluctuations can vary from cycle to cycle by a magnitude ($V-$band) or more. The rise times are $\sim1$ month while the decay times are $\sim2$ months, possibly correlated with the rapid heating timescale of the disk and photosphere as the mass accretion rate increases. The behavior has persisted over several decades at least.

Regular cadence, $g-$ and $r-$band imaging of V347 Aur obtained by the ZTF reveal that a small reflection nebula illuminated by the young stellar object also fluctuates in brightness, at times nearly fading completely. The reflection nebula appears to be the illuminated wall of an evacuated cavity, blown out of the L1438 molecular cloud presumably by a jet or outflow emerging from V347 Aur. V347 Aur is responsible for the brightness fluctuations of the nebula, which are produced by varying incident radiation from the embedded young stellar object.

Five epochs of high dispersion spectroscopy of V347 Aur obtained using HIRES on {\it Keck I} reveal two distinct spectral components, a quiescent M-dwarf absorption line spectrum and a bright emission line spectrum with a heavily veiled continuum. All spectra obtained exhibit strong H$\alpha$ emission as well as forbidden emission lines of [O I], [N II], and [S II], indicative magnetospheric accretion, jets and outflow activity. Low and high velocity components of \ion{Na}{1} $\lambda\lambda$5889, 5895, [O I] $\lambda\lambda$6600, 6363, and [S II] $\lambda\lambda$6717, 6731 emission are present with radial velocities averaging about $-$10 and $-$90 km s$^{-1}$, respectively.

Fourier cross-correlation analysis of the HIRES spectra reveal no detectable velocity offset between observations greater than $\sim$2 km s$^{-1}$. From the lack of radial velocity variability, we infer that V347 Aur is not a spectroscopic binary, i.e. no stellar mass companion is present, or the system is oriented such that the orbital plane is orthogonal to Earth's line of sight. Deep HST F814W imaging also finds no evidence for a bright companion close to V347 Aur. Although several more distant, faint sources are potentially identified, they cannot be responsible for the observed $\sim$160 day period.

We hypothesize that the semi-periodic brightness variations observed in V347 Aur arise from accretion instabilities induced by an undetected companion or a structure within the circumstellar disk that regulates gas flow. Similar to the EXor class of pre-main sequence variables, outbursts occur when gas builds up near the inner disk edge before crossing the co-rotation radius and flowing rapidly onto the stellar photosphere. The regular periodicity of the outbursts argues for a triggering mechanism that is linked to a specific orbital radius of $\sim$0.4 AU. V347 Aur along with V371 Ser may represent a class of eruptive pre-main sequence variables that experience periodic outbursts driven by accretion disk instabilities, possibly induced by a close companion.

V347 Aur is the only young stellar object definitively interacting with the L1438 molecular cloud. G. Herbig reported no H$\alpha$ emission sources associated with the cloud. We found from deep, near infrared imaging only a few sources that have colors consistent with circumstellar disk emission. It is very probable that V347 Aur is the most massive source forming within this molecular cloud core. 

On a larger scale, the L1438 molecular cloud is isolated from other star forming regions. The {\it Gaia}-measured parallax for V347 Aur implies a distance of 208.96$\pm$3.95 pc, well in the foreground of the Camelopardalis OB 1 association ($\sim$1 kpc) and other nearby open clusters. The northern-most aggregates of the Taurus-Auriga molecular cloud complex lie more than $\sim$20$^{\circ}$ south of L1438 and the nearby, similarly small, L1439 molecular cloud, implying a projected separation of $\sim$50 pc. While beyond the scope of the current investigation, it is suggested that V347 Aur is part of a loose stellar association that may be forming from these diffuse molecular clouds.

Future observational work is clearly needed including simultaneous multi-color photometric and spectroscopic observations of V347 Aur to correlate the photometric maxima with the prominent emission line spectrum. Additional high dispersion spectra in the near infrared where the contrast ratio between V347 Aur and a potential lower mass companion would be minimized would also be beneficial. High angular resolution, near infrared imaging of V347 Aur could be used to probe the inner disk ($<$0\farcs5) region for low mass companions or disk structure. Finally, millimeter continuum imaging of V347 Aur would aid in constraining circumstellar disk size and structure.

\acknowledgments We thank an anonymous referee for their helpful comments that improved this manuscript. SED is deeply indebted to G. H. Herbig who suggested that V347 Aur be targeted for future observations. Herbig provided SED with a short summary of his high dispersion spectra of V347 Aur obtained using HIRES on the {\it Keck I} telescope between 2003 and 2005. 

This paper has made use of the Digitized Sky Surveys, which were produced at the Space Telescope Science Institute under U.S. Government grant NAG W-2166, the SIMBAD database operated at CDS, Strasbourg, France, and the 2MASS, a joint project of the University of Massachusetts and the Infrared Processing and Analysis Center (IPAC)/California Institute of Technology, funded by NASA and the National Science Foundation. This publication makes use of data products from the Wide-field Infrared Survey Explorer, which is a joint project of the University of California, Los Angeles, and the Jet Propulsion Laboratory/California Institute of Technology, funded by the National Aeronautics and Space Administration. Some of the data presented herein were obtained at the W.M. Keck Observatory, which is operated as a scientific partnership among the California Institute of Technology, the University of California and the National Aeronautics and Space Administration. The Observatory was made possible by the generous financial support of the W.M. Keck Foundation. The Pan-STARRS1 Surveys (PS1) and the PS1 public science archive have been made possible through contributions by the Institute for Astronomy, the University of Hawaii, the Pan-STARRS Project Office, the Max-Planck Society and its 
participating institutes, the Max Planck Institute for Astronomy, Heidelberg and the Max Planck Institute for Extraterrestrial Physics, Garching, The Johns Hopkins University, Durham University, the University of Edinburgh, the Queen's University Belfast, the Harvard-Smithsonian Center for Astrophysics, the Las Cumbres Observatory Global Telescope Network Incorporated, the National Central University of Taiwan, the Space Telescope Science Institute, the National Aeronautics and Space Administration under Grant No. NNX08AR22G issued through the Planetary Science Division of the NASA Science Mission Directorate, the National Science Foundation Grant No. AST-1238877, the University of Maryland, Eotvos Lorand University (ELTE), the Los Alamos National Laboratory, and the Gordon and Betty Moore Foundation. ZTF is supported by the National Science Foundation and a collaboration including Caltech, IPAC, the Weizmann Institute for Science, the Oskar Klein Center at Stockholm University, the University of Maryland, the University of Washington, Deutsches Elektronen-Synchrotron and Humboldt University, Los Alamos National Laboratories, the TANGO Consortium of Taiwan, the University of Wisconsin at Milwaukee, and Lawrence Berkeley National Laboratories. Operations are conducted by COO, IPAC, and UW.

\newpage
\begin{deluxetable}{ccccccccccc}
\tabletypesize{\tiny}
\tablewidth{0pt}
\tablenum{1}
\tablecaption{Measured Equivalent Widths and Radial Velocities of Significant Emission and Absorption Features in V347 Aur}
\tablehead{
\colhead{Transition}  &   \multicolumn{2}{c}{2003 Dec 13 (A)}  & \multicolumn{2}{c}{2004 Sep 24 (B)} & \multicolumn{2}{c}{2004 Nov 21 (C)} & \multicolumn{2}{c}{2005 Nov 23 (D)} &  \multicolumn{2}{c}{2019 Nov 29 (E)}\\
                      &    v$_{\odot}$\tablenotemark{a}  &   EW    &   v$_{\odot}$\tablenotemark{a} &  EW    &    v$_{\odot}$\tablenotemark{a} &  EW    &  v$_{\odot}$\tablenotemark{a} &  EW &    v$_{\odot}$\tablenotemark{a} & EW  \\
                      &    km s$^{-1}$    &  \AA    &   km s$^{-1}$ &  \AA   &    km s$^{-1}$ &  \AA   &  km s$^{-1}$ &  \AA &    km s$^{-1}$ &  \AA    \\}
\startdata
He I $\lambda$5876    &  $+$14.7  &   $-$2.45         &  $+$11.5 &   $-$1.7      &   $+$13.6 &    $-$2.10      &  $+$13.0 &       $-$1.5   &  $+$15.7 &   $-$1.9   \\
Na I $\lambda$5889    &  $+$24.5  &   $-$0.65         &  $+$11.6 &   $-$2.0      &   $+$10.9 &    $-$2.65      &  $+$2.9  &       $-$1.6   &  $+$25.7 &   $-$3.0   \\
                   a  &  \tablenotemark{b} & ...      &  $-$39.0 &    ...        &   $-$38.6 &     ...         &  $-$39.8 &         ...    &  \tablenotemark{b}  & ... \\
                   b  &  \tablenotemark{b} & ...      &  $-$14.4 &    ...        &   $-$14.9 &     ...         &  $-$15.3 &         ...    &  \tablenotemark{b}  & ... \\
                  IS  &  $+$5.1   &     ...           &  $+$5.3  &    ...        &   $+$5.3  &     ...         &  $+$4.2  &         ...    &  $+$5.2  &     ...    \\
Na I $\lambda$5895    &	 $+$25.4  &    $-$0.4         &  $+$9.0  &  $-$1.7       &   $+$8.4  &    $-$2.33      &  $-$1.0  &       $-$1.6   &  $+$21.6 &  $-$2.4    \\ 
                   a  &  \tablenotemark{b} & ...      &  $-$39.0 &    ...        &   $-$39.4 &     ...         &  $-$40.3 &         ...    &  \tablenotemark{b}  & ... \\
                   b  &  \tablenotemark{b} & ...      &  $-$14.2 &    ...        &   $-$15.2 &     ...         &  $-$16.0 &         ...    &  \tablenotemark{b}  & ... \\
                  IS  &  $+$6.1   &   ...             &  $+$5.4  &    ...        &   $+$5.4  &     ...         &  $+$4.2  &         ...    &  $+$4.7  &    ...     \\ 
{[}O I{]}  $\lambda$6300    &  $-$44.2  &   $-$3.7          &  $-$52.7 &  $-$7.0       &   $-$48.2 &    $-$4.10      &  $-$52.2 &       $-$8.2   &  $-$45.8 &  $-$6.5 \\
                   HVC  &  $-$86.4  &   ...             &  $-$92.0 &  ...          &   $-$92.0 &     ...         &  $-$95.4 &         ...    &  $-$101.5 & ...     \\
                   LVC  &  $-$3.2   &   ...             &  $-$8.8  &  ...          &   $-$3.2  &     ...         &  $-$4.9  &         ...    &  $-$6.7 &    ...     \\
{[}O I{]} $\lambda$6363     &  $-$45.9  &   $-$1.1          &  $-$53.3 &  $-$1.7       &   $-$48.0 &    $-$1.19      &  $-$51.3 &       $-$2.0   &  $-$45.5 &  $-$1.5 \\
                   HVC  &  $-$86.2  &   ...             &  $-$90.8 &  ...          &   $-$91.2 &     ...         &  $-$95.1 &         ...    &  $-$101.0 &  ...    \\
                   LVC  &  $-$3.1   &   ...             &  $-$7.0  &  ...          &   $-$4.3  &     ...         &  $-$5.6  &         ...    &  $-7.9$ &     ...    \\
H$\alpha$             &  $+$47.0  &   $-$68.5         &  $+$18.5 & $-$65.6       &   $+$21.6 &    $-$85.7      &  $+$26.7 &       $-$72.1  &  $+$42.3 &  $-$82.0 \\
He I $\lambda$6678    &  $+$12.9  &   $-$1.2          &  $+$15.9 & $-$0.3        &   $+$7.5  &    $-$0.89      &  $+$12.8 &       $-$0.5   &  $+$12.9 &  $-$0.6 \\
Li I $\lambda$6708    &  $+$11.7 &  $+$0.1           &  $+$9.1  &  $+$0.4          &   $+$11.8 &    $+$0.3      &  $+$9.1  &       $+$0.4   &  $+$11.3 &  $+$0.3 \\
{[}S II{]} $\lambda$6716     &  $-$63.9  &     $-$0.2        &  $-$51.4 & $-$0.4        &   $-$60.1 &    $-$0.1      &  $-$50.6 &       $-$0.3   &  $-$77.6 &  $-$0.3 \\
                 HVC  &  $-$88.6  &      ...          & $-$99.5  & ...           &   $-$88.5 &     ...         &  $-$90.0 &         ...    &  $-$104.9 &  ...    \\
                 LVC  &  $-$14.9  &      ...          & $-$14.7  & ...           &   $-$13.9 &     ...         &  $-$13.7 &         ...    &  $-$13.9  &  ...    \\
{[}S II{]} $\lambda$6732     &  $-$61.7  &     $-$0.5        & $-$63.5  & $-$1.0        &   $-$64.3 &    $-$0.55      &  $-$61.9 &       $-$1.0   &  $-$69.0  & $-$0.8  \\
                 HVC  &  $-$87.8  &      ...          & $-$89.3  & ...           &   $-$89.2 &     ...         &  $-$86.2 &         ...    &  $-$99.4  &  ...    \\
                 LVC  &  $-$16.3  &      ...          & $-$17.8  & ...           &   $-$16.3 &     ...         &  $-$17.4 &         ...    &  $-$17.0  &  ...    \\
\hline
\enddata
\tablenotetext{a}{The heliocentric radial velocities presented are based purely on the applied barycentric corrections and wavelength calibration procedure. Instrumental errors are anticipated to be 1-2 km s$^{-1}$.}
\tablenotetext{b}{In the 2003 Dec 13 and 2019 Nov 29 spectra, the a and b components were blended into a single absorption feature.}
\end{deluxetable}

\end{document}